\documentclass[twocolumn]{aastex62}

\usepackage{comment}

\newcommand{\Msun}{{\rm M_{\odot}}}

\newcommand{\kmps}{\, {\rm km \, s^{-1}}}

\newcommand{\Ratio}{{R_{21/10}}}

\shorttitle{CO $J$=2-1/1-0 Variations in M83}
\shortauthors{Koda et al.}

\begin{document}

\title{Systematic Variations of CO $J$=2-1/1-0 Ratio and Their Implications in The Nearby Barred Spiral Galaxy M83}


\correspondingauthor{Jin Koda}
\email{jin.koda@stonybrook.edu}

\author{Jin Koda}
\affiliation{Department of Physics and Astronomy, Stony Brook University, Stony Brook, NY 11794-3800, USA}
\affiliation{Amanogawa Galaxy Astronomy Research Center, Kagoshima University, 890-0065, Kagoshima, Japan\footnote{visiting}}

\author{Tsuyoshi Sawada}
\affiliation{NAOJ Chile, National Astronomical Observatory of Japan, Alonso de C\'ordova 3788, Office 61B, Vitacura, Santiago 763 0492, Chile}
\affiliation{Joint ALMA Observatory, Alonso de C\'ordova 3107, Vitacura, Santiago 763 0355, Chile}

\author{Kazushi Sakamoto}
\affiliation{Academia Sinica, Institute of Astronomy and Astrophysics, Taipei 10617, Taiwan}

\author{Akihiko Hirota}
\affiliation{NAOJ Chile, National Astronomical Observatory of Japan, Alonso de C\'ordova 3788, Office 61B, Vitacura, Santiago 763 0492, Chile}
\affiliation{Joint ALMA Observatory, Alonso de C\'ordova 3107, Vitacura, Santiago 763 0355, Chile}

\author{Fumi Egusa}
\affiliation{Institute of Astronomy, Graduate School of Science, The University of Tokyo, 2-21-1 Osawa, Mitaka, Tokyo 181-0015, Japan}

\author{Samuel Boissier}
\affiliation{Aix Marseille Univ., CNRS, CNES, Laboratoire d'Astrophysique de Marseille, Marseille, France}

\author{Daniela Calzetti}
\affiliation{Department of Astronomy, University of Massachusetts, Amherst, MA 01002, USA}

\author{Jennifer Donovan Meyer}
\affiliation{National Radio Astronomy Observatory, 520 Edgemont Road, Charlottesville, VA 22903-2475, USA}

\author{Bruce G. Elmegreen}
\affiliation{IBM Research Division, T. J. Watson Research Center, 1101 Kitchawan Road, Yorktown Heights, NY 10598, USA}

\author{Armando Gil de Paz}
\affiliation{Departamento de F\'{\i}sica de la Tierra y Astrof\'{\i}sica, Universidad Complutense de Madrid, Plaza Ciencias 1, Madrid E-28040, Spain \& Instituto de F\'{\i}sica de Part\'{\i}culas y del Cosmos (IPARCOS)}

\author{Nanase Harada}
\affiliation{Academia Sinica, Institute of Astronomy and Astrophysics, Taipei 10617, Taiwan}

\author{Luis C. Ho}
\affiliation{The Kavli Institute for Astronomy and Astrophysics, Peking University, 5 Yiheyuan Road, Haidian District, Beijing, 100871, China}
\affiliation{Department of Astronomy, Peking University, 5 Yiheyuan Road, Haidian District, Beijing, 100871, China}

\author{Masato I.N. Kobayashi}
\affiliation{Department of Earth and Space Science, Graduate School of Science, Osaka University, Osaka 560-0043, Japan}

\author{Nario Kuno}
\affiliation{Department of Physics, Graduate School of Pure and Applied Sciences, University of Tsukuba, 1-1-1 Ten-nodai, Tsukuba, Ibaraki 305-
8577, Japan}
\affiliation{Tomonaga Center for the History of the Universe, University of Tsukuba, 1-1-1, Ten-nodai, Tsukuba, Ibaraki 305-8571, Japan}

\author{Sergio Mart\'in}
\affiliation{European Southern Observatory, Alonso de C\'ordova, 3107, Vitacura, Santiago 763-0355, Chile}
\affiliation{Joint ALMA Observatory, Alonso de C\'ordova 3107, Vitacura, Santiago 763 0355, Chile}

\author{Kazuyuki Muraoka}
\affil{Department of Physical Science, Graduate School of Science, Osaka Prefecture University, 1-1 Gakuen-cho, Naka-ku, Sakai, Osaka 599-8531, Japan}

\author{Kouichiro Nakanishi}
\affiliation{National Astronomical Observatory of Japan, Mitaka, Tokyo 181-8588, Japan}
\affiliation{The Graduate University for Advanced Studies, SOKENDAI, Mitaka, Tokyo 181-8588, Japan}

\author{Nick Scoville}
\affiliation{California Institute of Technology, MC 249-17, 1200 East California Boulevard, Pasadena, CA 91125, USA}

 \author{Mark Seibert}
 \affiliation{Observatories of the Carnegie Institution for Science, Pasadena, CA 91101}

\author{Catherine Vlahakis}
\affiliation{National Radio Astronomy Observatory, 520 Edgemont Road, Charlottesville, VA 22903-2475, USA}

\author{Yoshimasa Watanabe}
\affiliation{College of Engineering, Nihon University, 1 Nakagawara, Tokusada, Tamuramachi, Koriyama, Fukushima 963-8642, Japan}

\begin{abstract}
We present spatial variations of the CO $J$=2-1/1-0 line ratio ($\Ratio$) in the barred spiral galaxy M83
using Total Power array (single dish telescopes) data from the Atacama Large Millimeter/submillimeter Array (ALMA).
While the intensities of these two lines correlate tightly,
$\Ratio$ varies over the disk, with a disk average ratio of 0.69, and shows the galactic center and a two-arm spiral pattern.
It is high ($\gtrsim0.7$) in regions of high molecular gas surface density ($\Sigma_{\rm mol}$),
but ranges from low to high ratios in regions of low $\Sigma_{\rm mol}$.
The ratio correlates well with the spatial distributions and intensities of far-ultraviolet (FUV) and infrared (IR) emissions,
with FUV being the best correlated.
It also correlates better with the ratio of specific intensities at 70 and 350$\mu$m, a proxy for dust temperature, than with the IR intensities.
Taken together, these results suggest either a direct or indirect link between the dust heating
by the interstellar radiation field (ISRF) and the condition of GMCs,
even though no efficient mechanism is known for a thermal coupling of dust and bulk gas in GMCs.
We speculate that the large spread of  $\Ratio$ in low $\Sigma_{\rm mol}$ regions,
mostly at the downstream sides of spiral arms,
may be due to the evolution of massive stars after spiral arm passage.
Having in a late phase escaped from the spiral arms and their parental clouds,
they may contribute to the dust heating by FUV and gas heating by cosmic rays produced by supernovae.
\end{abstract}


\section{Introduction} \label{sec:intro}

The $J$=1-0 line transition of carbon monoxide, CO(1-0),
has been the yardstick for observations and calibrations of the molecular gas in the Milky Way (MW) and
nearby galaxies \citep[see ][for review]{Scoville:1987vo, Dame:1987aa, Fukui:2010lr, Heyer:2015qy}.
Recently, this fundamental transition is being replaced by the higher excitation transition CO(2-1)
on the assumption of a constant CO 2-1/1-0 line ratio \citep[$\Ratio$; e.g., ][]{Leroy:2009zv, Sun:2018aa, Saintonge:2018aa}.
Observations in CO(2-1) require much less time than those in CO(1-0) 
to achieve the same mass sensitivity especially at the ALMA site \citep{Sakamoto:2008aa, Watson:2017aa}.
Many nearby galaxy projects with ALMA employ CO(2-1) as an alternative to CO(1-0) to trace the bulk molecular gas.

The notion of a constant $\Ratio$ arose from analyses of past single-dish data on
nearby galaxies \citep[e.g., ][]{Bigiel:2008aa, Sandstrom:2013vn},
with caveats \citep{Leroy:2009zv}.
The faint CO emission in the interarm regions was often undetected,
and most of those analyses were limited to radial profiles
after azimuthal averaging (hence washing out arm-interarm variations).
Measurements of $\Ratio$ often suffered from calibration difficulties \citep{Koda:2012lr}.
For example, no obvious variation was found in M51 with earlier data \citep{Garcia-Burillo:1993ys},
but systematic variations between the spiral arms and interarm regions were found later,
primarily due to improved observational instruments and techniques \citep{Koda:2012lr, Vlahakis:2013aa}.

It is known that $\Ratio$ is an important diagnostic tracer of the physical conditions of molecular gas.
In the MW, $\Ratio$ changes systematically from 1.0-1.2 to 0.3-0.4 between spiral arms
and interarm regions,
from the galaxy center to the outer disk, and between star-forming and dormant giant molecular clouds
\citep[GMCs; ][]{Sakamoto:1994lr, Sakamoto:1997ys, Oka:1996aa, Hasegawa:1997lr, Falgarone:1998aa, Seta:1998aa, Sawada:2001lr, Yoda:2010rf, Nishimura:2015aa}.
These variations in the MW and M51 can be interpreted as changes of a factor of 2-3 in temperature and/or density,
according to the non-LTE calculations \citep{Scoville:1974yu, Goldreich:1974jh, Koda:2012lr}.
Besides these two galaxies, analyses of $\Ratio$ with well-calibrated data still remain rare
even with single dish telescopes.
It is urgent to build up such accurate analyses given the growing amount of CO(2-1) observations
of nearby galaxies.

Here we present another case, the barred spiral galaxy M83 at a distance of 4.5 Mpc \citep{Thim:2003aa},
using new single-dish data from ALMA.
In this galaxy, \citet{Crosthwaite:2002yu} found an elevated $\Ratio$ in the interarm regions,
contrary to the results in the MW and M51.
The ratio appeared so high ($>1$) that it potentially indicated that optically-thin CO emission
dominates in the interarm regions and overshadows the emission from GMCs.
\citet{Lundgren:2004aa} also found a similar qualitative trend: an elevated, but lower ($<1$), $\Ratio$ in the interarm regions.
This ratio can be explained by the optically-thick molecular gas within GMCs.
We show that the enhanced $\Ratio$ occurs at the downstream sides of the spiral arms.
In the interarm regions farther away from the arms,
$\Ratio$ becomes lower and is consistent with that observed in the MW and M51.
The new example of $\Ratio$ variations emphasizes the importance of $\Ratio$
as a prime diagnostic tool of the physical condition of bulk molecular gas in galaxies.

\section{Observations and Data Reduction} \label{sec:obs}

M83 was observed with the Total Power (TP) array of ALMA in CO(1-0) and CO(2-1).
After the data reduction described below, we analyze the data at the full-width half maximum (FWHM) beam size of
the CO(1-0) data, $56.6\arcsec$ ($\sim1.2$ kpc).
Despite its lower spatial resolution, the analysis of single-dish data is an important first step
for a solid confirmation of the variations of $\Ratio$,
since interferometer data are susceptible to additional noise introduced in the imaging process.
The data were reduced using the Common Astronomy Software Applications package \citep[CASA; ][]{McMullin:2007aa}.
The calibration was performed in the standard way as for the ALMA data reduction pipeline,
with a more careful calibration of the relative flux scales among execution blocks (EBs; see below).

\subsection{CO(1-0)}\label{sec:co10}

The CO(1-0) observations mapped a $11.7\arcmin \times 11.7\arcmin$ area
with the On-The-Fly (OTF) mapping technique along the RA and Dec directions.
We obtained a total of 125 EBs, each of which, with a few exceptions, observed the whole area.
We rejected 5 EBs due to bad weather (unreasonably low intensity),
spurious pointing corrections (blurred map appearance),
and relatively large flux errors with respect to the other EBs (deviations greater than a few \%).
The statistics of the 120 EBs are:
the average number of TP antennas was 3.71,
the total observing and on-source times were 110.1 h and 60.4 h, respectively, and
the system temperature
$T_{\rm sys}$ was 99-129~K at the 10-90 percentiles.
The ACA correlator was used to obtain spectra with band and channel widths of 250~MHz and 61~kHz.

Individual spectra were calibrated in the standard way.
After the calibration they were re-sampled on a grid of pixel size $5.62\arcsec$
using the prolate spheroidal function with a size of 6 pixels \citep{Schwab:1980aa, Schwab:1984}.
The effective FWHM beam size after this regridding/smoothing is $56.6\arcsec$.
We generated separate data cubes for the 120 EBs, calculated the flux ratios and errors
of all of their pairs, and solved for relative flux scales by inverting the design matrix.
The derived scaling coefficients have a small scatter of only 1.3\%.
We applied these coefficients to co-add all the spectra into two data cubes
for the RA and DEC scans.
Spectral baselines were subtracted with straight lines.
The two were combined with the \citet{Emerson:1988hz} method.
We converted the antenna temperature $T_{a}^*$ into the main beam temperature $T_{\rm mb}$
using a main beam efficiency of 0.856 calculated from the Jy/K parameter from the observatory.
The final cube has 3,900 channels and an root-mean-square noise of $\sim 6.2$ mK in $T_{\rm mb}$
in a velocity channel width of $0.159\kmps$.

\subsection{CO(2-1)}\label{sec:co21}

The CO(2-1) observations covered approximately a $8.4\arcmin \times8.4\arcmin$ region
with a position angle of about $52\deg$.
The region was split into 9 rectangular regions,
each of which was observed separately.
About 30\% the area of each region overlaps with that of an adjacent region.
The total number of EBs was 160.
The average number of antennas per EB was 3.57.
The total observing and on-source times were 132.8 h and 66.4h.
The $T_{\rm sys}$ was 59-89~K at the 10-90 percentiles.
The band and channel widths were 2~GHz and 977~kHz or 1~GHz and 488~kHz.

All the data were reduced in the standard way.
We generated 160 separate data cubes for all the EBs,
calculated the flux ratios and errors of all their pairs when their spatial coverages overlapped,
and solved for the relative flux scaling coefficients.
The coefficients have a small scatter of 3.1\%,
and we made a correction for this before co-adding the data.
We used a main beam efficiency of 0.836.
The final data cube has 488 channels and
an RMS noise of $\sim 2.0$ mK in $T_{\rm mb}$
with a pixel scale of $2.81\arcsec$ and velocity channel width of $1.276\kmps$.

\subsection{The CO 2-1/1-0 Ratio ($\Ratio$)}\label{sec:ratio}

\begin{figure*}[h]
\plotone{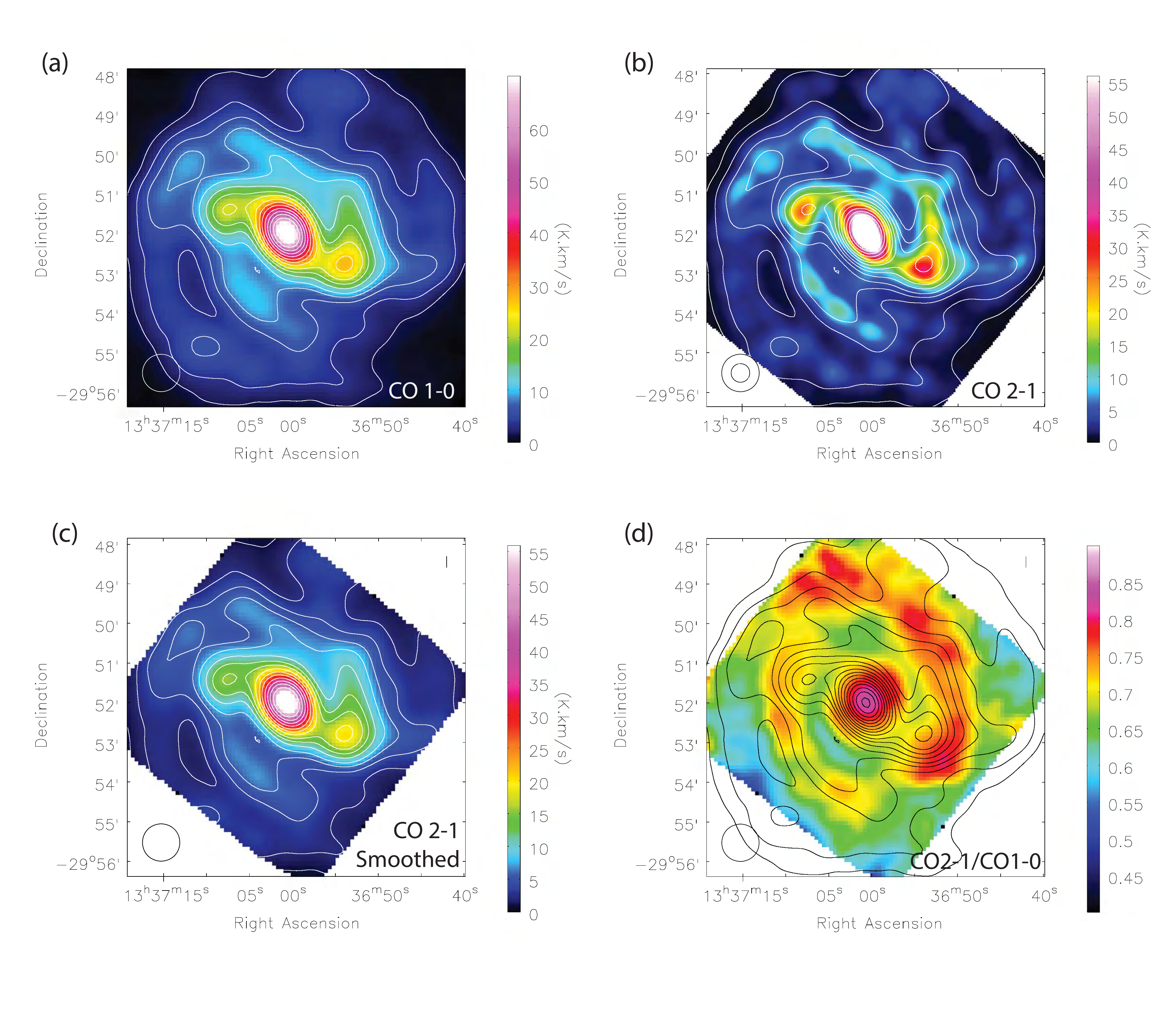}
\caption{Maps of M83 in
(a) CO(1-0),
(b) CO(2-1),
(c) CO(2-1) smoothed to the CO(1-0) resolution, and
(d) CO 2-1/1-0 ratio ($\Ratio$).
The contours are of CO(1-0) at 1.5, 3, 5, 7, 9, 12, 15, 18, 22, 26, 30, 35, 40, 45, 51, 57, and 63 $\, \rm K \cdot \kmps$.
The beam sizes of the original CO(1-0) and CO(2-1) maps (a \& b) are indicated by circles ($56.6\arcsec$ and $28.3\arcsec$, respectively).
For panels (c) and (d), we trimmed the edges where the blank data outside the CO(2-1) field-of-view affect the smoothing.
\label{fig:maps}}
\end{figure*}

The CO(1-0) and CO(2-1) data cubes were integrated along velocity without any clip/mask,
to produce integrated intensity maps, $I_{\rm CO1-0}$ and $I_{\rm CO2-1}$
 (Figures \ref{fig:maps}a, b).
To match the spatial resolutions, the CO(2-1) map was smoothed by a sequence of
a deconvolution with the CO(2-1) beam and convolution with the CO(1-0) beam.
The beams here are generated by a convolution of the native telescope beam,
approximated with a gaussian (a FWHM of $58.3\arcsec$ at 100~GHz),
and the spheroidal function.
The smoothed map was regridded to the pixel scale of the CO(1-0) map.
The $1\sigma$ noise is about 61 and 14 $\rm \rm mK \cdot km/s$ in $I_{\rm CO1-0}$ and $I_{\rm CO2-1}$,
respectively.
The lowest CO(1-0) contour in Figure 1 is at about 25$\sigma$ significance,
and the same locations are at about 70$\sigma$ significance in CO(2-1).

The CO(1-0) beam is larger than the CO(2-1) beam, and the edge of the smoothed map suffers from
the absence of data beyond the field coverage.
To quantify this effect, we applied the above smoothing procedure to an image with
a uniform illumination over the field of view (i.e., the pixel value was set to 1 within the field
of view and 0 outside). 
We used the pixels with a value greater than 0.99 after the smoothing.
This reduced the field-of-view to about $7.0\arcmin \times 7.0 \arcmin$.
The final CO(2-1) map is presented in Figure \ref{fig:maps}c.
The $\Ratio$ is calculated as $I_{\rm CO2-1}/I_{\rm CO1-0}$ (Figure \ref{fig:maps}d).
The random error in $\Ratio$ is about 4\% at the lowest $I_{\rm CO1-0}$ contour or less at higher $I_{\rm CO1-0}$.

These maps suffer from the systematic errors in absolute flux calibrations (about 5\% according to the observatory),
which however does not affect the relative variations of $I_{\rm CO2-1}$, $I_{\rm CO1-0}$, and $\Ratio$ within the maps.
This 5\% systematic uncertainty mainly comes from the uncertainty in the models of primary flux calibrators.
It is likely to affect both $I_{\rm CO2-1}$ and $I_{\rm CO1-0}$ in a similar manner and, to an extent, cancel out in $\Ratio$.

The sampling of these maps is redundant with a pixel scale of $5.62\arcsec$
for a beam size of $56.6\arcsec$.
We use only every 5th pixel when they are correlated with other data (Sec. \ref{sec:results}).

\subsection{Ancillary Data}\label{sec:ancillary}

For comparisons, we obtained ancillary data from the archives:
specific intensity (brightness) maps in GALEX FUV (1516$\AA$)  \citep[$I_{\rm FUV}$; ][]{Gil-de-Paz:2007lj},
in Spitzer 24$\mu$m \citep[$I_{\rm 24\mu m}$; ][]{Bendo:2012aa},
and in Herschel 70 and 350$\mu$m \citep[$I_{\rm 70\mu m}$ and  $I_{\rm 350\mu m}$; ][]{Bendo:2012ab}.
Their native resolutions are 4.2, 5.7, 5.6, and 24.2$\arcsec$, respectively.
For comparisons with the CO data,
they are smoothed to a Gaussian beam
and convolved with Schwab's spheroidal function,
resulting in a final beam size of $56.6\arcsec$.
The data are then regridded to the CO data grid.

\section{Results} \label{sec:results}

Figure \ref{fig:maps} shows that the CO line ratio varies spatially
between the galactic center and disk, and between the spiral arms and interarm regions.
For comparison, the CO(2-1) map (Figure \ref{fig:maps}b) shows the locations of the molecular spiral arms.
Both the average and median of $\Ratio$ in this region are 0.69.
The observed range of $\sim$0.5-0.8 likely represents the upper and lower limits due to the low resolution (see discussions below).
Still, it appears that $\Ratio$ varies from $<0.7$ in the interarm regions to $>0.7$ around the spiral arms
-- or more precisely, at the convex,
presumably downstream
\footnote{HII regions appear preferentially on the convex sides \citep{Poetrodjojo:2019aa}, and hence we assume they are the downstream sides.},
sides of the spiral arms.
This is consistent with the variations in the MW \citep{Sakamoto:1997ys} and in M51 \citep{Koda:2012lr}.
The north-western spiral arm shows a higher ratio than the south-eastern arm.
Along the arms, $\Ratio$ is approximately constant and does not show clear radial trends.
$\Ratio$ is high$>0.8$ in the galactic center, as is also observed in the MW center \citep{Oka:1996aa, Sawada:2001lr}.
It stays $\sim 0.7$ along the bar.

Despite the low resolution, this analysis of single dish data is an important first confirmation of the variations,
since the flux calibration is fairly consistent over the galactic disk.
It is still important to keep in mind that in Figure \ref{fig:maps} the galactic structures,
such as the spiral arms and interam regions, are somewhat smeared due to the large beam size.
The range of $\Ratio$, if it is observed at a higher resolution, would be wider,
while the mean and median $\Ratio$ are less likely affected.
The typical range of $\Ratio$ observed in the MW and in M51 at higher linear resolutions
is 0.4 - 1.2, with only rare instances of $>1$.
It is conceivable that the intrinsic range in M83 is similar to those in the MW and M51,
but is smoothed to the observed range.

A non-LTE model suggests that
$\Ratio$ depends primarily on the H$_2$ volume density $n_{\rm H_2}$ and kinetic temperature $T_{\rm k}$ for collisional excitation,
and, on the CO column density $N_{\rm CO}$ per velocity (or optical depth) for photon trapping and radiative transfer,
although this last factor is typically constrained within a relatively narrow range
\citep{Scoville:1974yu, Goldreich:1974jh, Solomon:1987pr, Koda:2012lr}.
Obviously, our spatial resolution element (or beam; $\sim$1.2 kpc) includes multiple GMCs (see Figure \ref{fig:maps}).
Even the lowest contour corresponds to a molecular gas mass of $\rm M_{\rm mol}\sim 8\times10^6\Msun$
in one beam \citep[using the conversion factor from ][ to convert
$I_{\rm CO1-0}$ into $\rm M_{\rm mol}$ or molecular gas surface density $\Sigma_{\rm mol}$]{Bolatto:2013ys}.
This $\rm M_{\rm mol}$ is about 20 GMCs if all the emission is from GMCs analogous to
a typical Galactic GMC \citep[$4\times 10^5\Msun$; ][]{Scoville:1987vo}.
The $n_{\rm H_2}$ and $T_{\rm k}$ are the parameters \textit{within} the GMCs.
If we assume that all GMCs in each beam have the same average conditions,
the factor of $\sim2$ change in $\Ratio$ roughly corresponds to changes of a factor of $\sim$2-3
in $n_{\rm H_2}$ and/or $T_{\rm k}$ in the GMCs \citep{Koda:2012lr}.

\begin{figure*}[h]
\plotone{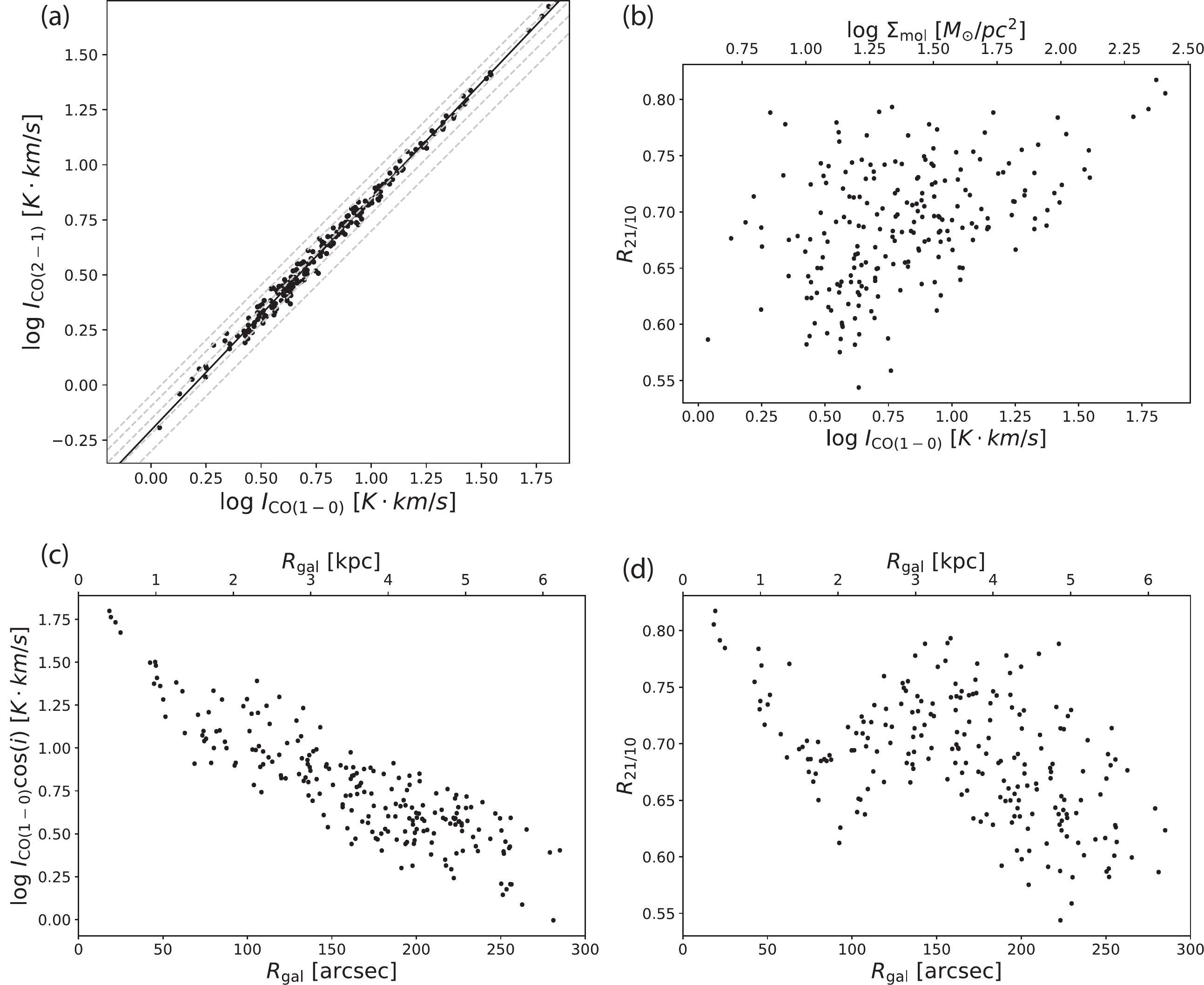}
\caption{
(a) $I_{\rm CO1-0}$ vs $I_{\rm CO2-1}$.
Dashed lines represent $\Ratio$=0.9, 0.8, 0.7, 0.6, and 0.5 (top to bottom).
The solid line is a fit.
(b) $\Ratio$ vs $I_{\rm CO1-0}$ ($\Sigma_{\rm mol}$).
Radial profiles of
(c) $I_{\rm CO1-0}\cos i$ with inclination correction, and (d)  $\Ratio$.
The radius $R_{\rm gal}$ is calculated
from the galactic center (RA, DEC) = (13:37:00.8, -29:51:58)
using the inclination ($i=25\deg$) and position angle (PA=$226\deg$)
from \citet{Crosthwaite:2002yu}.
The random error in $\Ratio$ is $<4$\%.
The maps in Figure \ref{fig:maps} are spatially oversampled,
and we use only every 5th pixel (about half-beam sampling) for these scatter plots.
\label{fig:corr1}}
\end{figure*}

$I_{\rm CO1-0}$ and $I_{\rm CO2-1}$ are tightly correlated (Figure \ref{fig:corr1}a),
while their ratio $\Ratio$ also changes with $I_{\rm CO1-0}$ (or $\Sigma_{\rm mol}$; Figure \ref{fig:corr1}b).
The bottom-right part of this figure is empty as if the region below a diagonal line were avoided.
The $I_{\rm CO1-0}$ is defined over a large beam, and hence, $\Sigma_{\rm mol}$ represents
the average surface density in the environment of unresolved GMCs within the beam.
In this plot, the $\Ratio$ is always high ($\gtrsim 0.7$) in regions of high $\Sigma_{\rm mol}$ (or $I_{\rm CO1-0}$).
The GMCs within the beam have higher $n_{\rm H_2}$ and/or $T_{\rm k}$ on average
when their environment is crowded with GMCs.
On the other hand, the ratio is spread over a relatively wide range in regions of low $\Sigma_{\rm mol}$.
Even though the average environmental surface density is low,
$n_{\rm H_2}$ and/or $T_{\rm k}$ within the underlying GMCs can vary from low to high,
resulting in the spread in $\Ratio$.

The radial profiles of $I_{\rm CO1-0}$ and $\Ratio$ are shown in Figure \ref{fig:corr1}c and d.
An inclination correction is applied to $I_{\rm CO1-0}$.
$I_{\rm CO1-0}$ (and $\Sigma_{\rm mol}$) decays exponentially with a scale length of
$\sim 94\arcsec$ ($\sim$2.0~kpc).
Note that the profile of $I_{\rm CO2-1}$ is similar with a scale length of $\sim 88\arcsec$ (1.9~kpc).
$\Ratio$ peaks at the galactic center ($\gtrsim 0.8$), decreases through the radius range of the bar ($\sim 0.65$),
increases again due to the spiral arms ($\sim 0.75$), and decreases toward the outskirts ($\lesssim 0.60$).
Note again that these are the ratios averaged over a 1.2~kpc beam.

\begin{figure*}[h]
\plotone{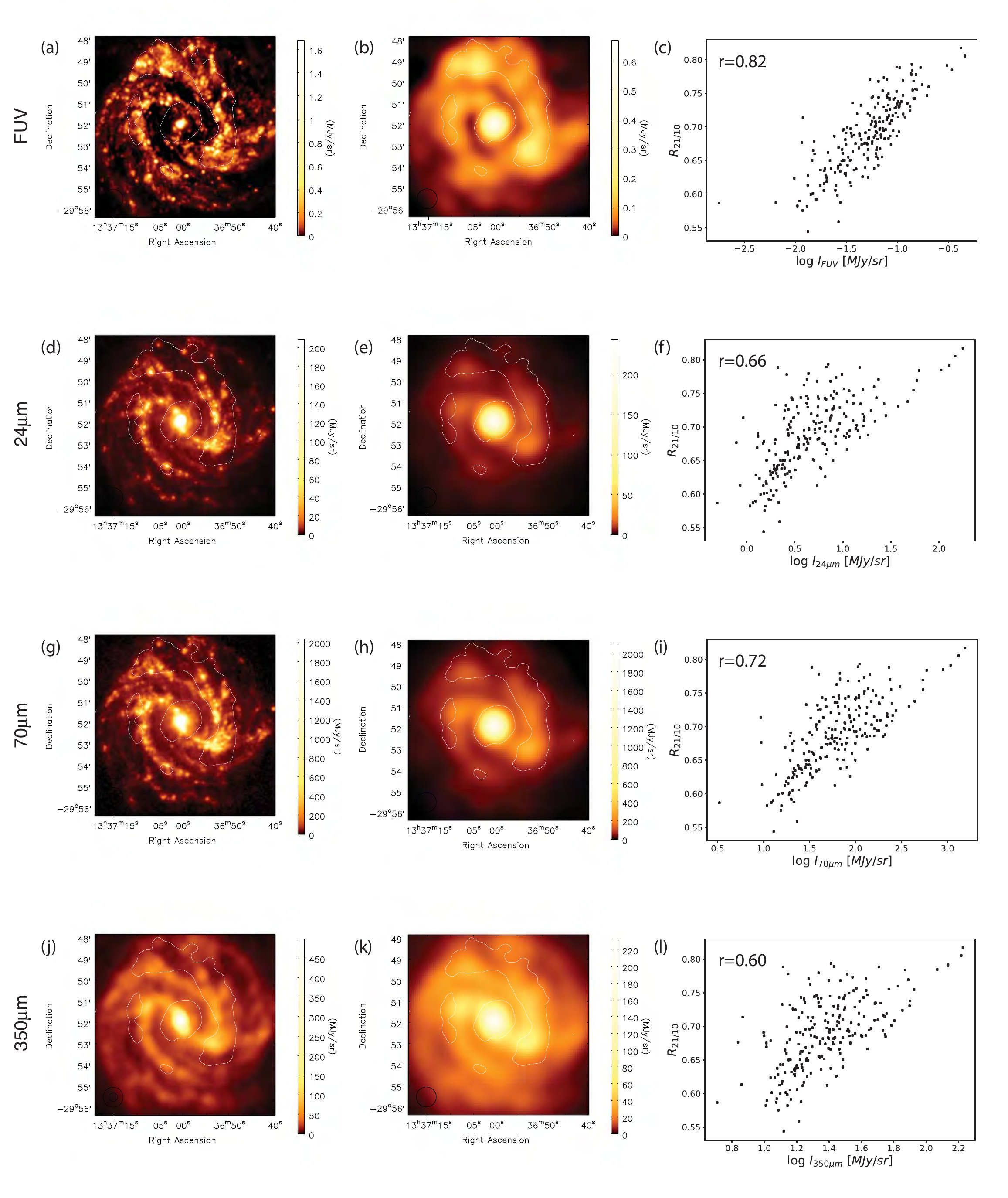}
\caption{
Rows 1-4: GALEX FUV, Spitzer $24\mu$m, Herschel 70$\mu$m and 350$\mu$m data.
Column 1: images at the data's native resolutions of 4.2, 5.7, 5.6, and 24.2$\arcsec$ from top to bottom.
Contours are at $\Ratio=0.72$ to enclose the regions of high $\Ratio$.
Column 2: images smoothed to the CO(1-0) resolution of 56.6$\arcsec$. 
Column 3: correlations between $\Ratio$ and each data in specific intensity.
The correlation coefficients $r$ are shown.
\label{fig:ancillary}}
\end{figure*}

Figure \ref{fig:ancillary} compares $\Ratio$ with the FUV and IR data.
The contours in the left and middle columns indicate regions of high $\Ratio$.
The FUV image shows a tighter spatial correlation with high $\Ratio$ than the IR images
even though the FUV image suffers more from extinction.
In particular, FUV and $\Ratio$ remain high at the downstream sides of the spiral arms.
The right column shows correlations between $\Ratio$ and their specific intensities.
The correlation coefficient $r$ is shown in each plot (as well as in Table \ref{tab:fits},
which also lists the gradients and intercepts from linear regression).
FUV presents the tightest correlation here as well.
$\Ratio$ also correlates with the IR emissions.
It is high in the regions of high $I_{\rm 24\mu m}$, $I_{\rm 70\mu m}$, and  $I_{\rm 350\mu m}$,
but shows relatively large spreads at low specific intensities
[this is similar to what is seen with $I_{\rm CO1-0}$  ($\Sigma_{\rm mol}$)].
The regions of high $\Ratio$ at low IR intensities are a major contributor
to the spreads in their correlations (i.e., top-left quartiles in Figure \ref{fig:ancillary}f, i, l).
Those regions tend to have low $I_{\rm CO1-0}$  and
contribute also to the spreads in the correlation of $\Ratio$ and $I_{\rm CO1-0}$ (Figure \ref{fig:corr1}b).
On the other hand, they have high FUV intensities and lie on the tighter correlation of $\Ratio$ and FUV  (Figure \ref{fig:ancillary}c).
They tend to appear on the convex, downstream sides of the spiral arms,
and might correspond to the ``interarm" gas with an elevated $\Ratio$
found by \citet{Crosthwaite:2002yu} and \citet{Lundgren:2004aa}.

\begin{deluxetable}{ccccc}[b!]
\tablecaption{Results of Linear Regression \label{tab:fits}}
\tablecolumns{5}
\tablewidth{0pt}
\tablehead{
\colhead{$x$} & 
\colhead{$y$} & 
\colhead{$\alpha$} & 
\colhead{$\beta$} &
\colhead{$r$}
}
\startdata
        $\log I_{\rm CO1-0}$ &         $\log I_{\rm CO2-1}$ & 1.05 & -0.205 &  1.00 \\          
        $\log I_{\rm CO1-0}$ &         $\Ratio$                    &  0.219 &  0.511 &  0.45 \\
        $\log I_{\rm CO2-1}$ &         $\Ratio$                    &  0.190 &  0.565 &  0.53 \\
        $\log I_{\rm FUV}$ &             $\Ratio$                    & 0.161 &  0.897 &  0.82 \\ 
        $\log I_{\rm 24\mu m}$ &      $\Ratio$                    & 0.137 &  0.596 &  0.66 \\ 
        $\log I_{\rm 70\mu m}$ &      $\Ratio$                    & 0.134 &  0.448 &  0.72 \\ 
        $\log I_{\rm 350\mu m}$ &    $\Ratio$                    & 0.242 &  0.355 &  0.60 \\
    $\log I_{\rm 70\mu m}/I_{\rm 350\mu m}$ &  $\Ratio$& 0.292 & 0.569 &  0.83 \\ 
\enddata
\tablecomments{The gradient ($\alpha$) and intercept ($\beta$)
from the ordinary least squares bisector fitting \citep{Isobe:1990aa},
and the correlation coefficient ($r$).}
\end{deluxetable}

\begin{figure*}[h]
\plotone{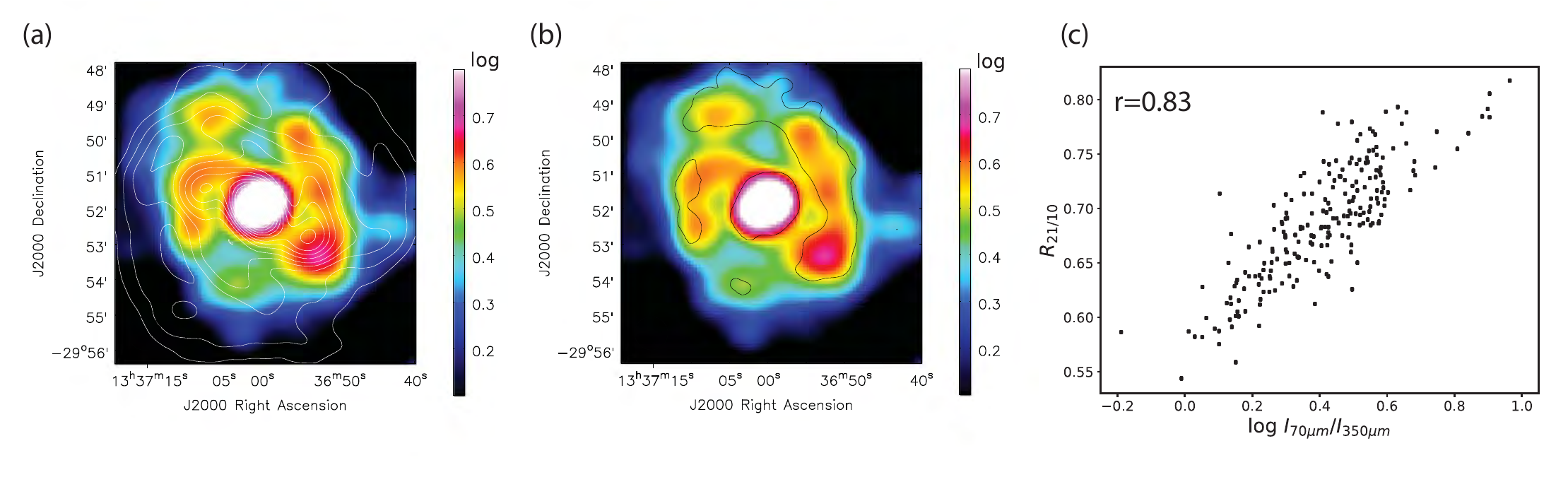}
\caption{
Comparisons with $\log (I_{70\mu m}/I_{350\mu m})$ (color), a proxy of dust temperature.
The contours are (a) the CO(1-0) integrated intensity, and (b) $\Ratio=0.72$.
(c) $\Ratio$ vs $\log (I_{70\mu m}/I_{350\mu m})$.
Note this dust color map is similar to a $T_{\rm d}$ map from an SED fitting by \citet{Foyle:2011lr}.
\label{fig:f70f350}}
\end{figure*}

FUV is expected to heat up dust grains, and in fact, $\Ratio$ also correlates well
with $I_{70\mu m}/I_{350\mu m}$ (dust color, a proxy of dust temperature $T_{\rm d}$; Figure  \ref{fig:f70f350}).
Here we choose $70\mu$m and $350\mu$m to cover the expected range of $T_{\rm d}$
of around 10-40 K (using the peak of blackbody spectrum as a rough guideline).
Theoretically, no efficient coupling mechanism between $T_{\rm d}$ and gas temperature $T_{\rm k}$ is
identified yet \citep{Goldreich:1974jh, Scoville:1976qy, Goldsmith:1978zh} at the average density of Galactic GMCs \citep[$\sim 300 \,\rm cm^{-3}$; ][]{Scoville:1987vo}.
However, this correlation suggests a link, either directly or indirectly, between the dust heating
by the interstellar radiation field (ISRF) and the condition of GMCs even after the passages of spiral arms.

\section{Discussion} \label{sec:discussion}
We showed a systematic trend that $\Ratio$ is elevated in regions of high $\Sigma_{\rm mol}$
(when they are averaged over a 1.2~kpc area), while its spread becomes large in
environments of low $\Sigma_{\rm mol}$.
In particular, the high $\Ratio$ at low $\Sigma_{\rm mol}$ is not immediately expected and requires additional explanation.
Such a condition is found mostly around the radial range of the spiral arms (Figure \ref{fig:corr1}d).
This high $\Ratio$ shows good quantitative and positional correlations with high $I_{\rm FUV}$
and $T_{\rm d}$ everywhere, including the downstream side of the spiral arms, better than with the IR intensities.

A possible explanation may be a potential direct thermal coupling between the dust, heated by UV photons, and the gas.
Although no efficient mechanism has been identified \citep{Goldsmith:1978zh},
a potential mechanism might be a coupling through H$_2$O molecules -
they may absorb dust radiation and heat up H$_2$ by collisional de-excitation \citep{Scoville:1976qy}.
Other possibilities are more indirect.
GMCs may, in some way, remain dense and/or warm after spiral arm passage and maintain a high $\Ratio$
on a timescale of $\sim100$ Myr, which is about the lifetime of B stars, a source of bright FUV.

We could also speculate that the observed trend is related to the evolution of massive stars after arm passage.
The massive stars are initially obscured in their parental clouds, contributing to
the IR intensities through dust heating.
They later escape the parental clouds and enhance the ISRF in-between GMCs (thus, high unobscured $I_{\rm FUV}$)
and probably the cosmic ray (CR) flux through subsequent supernovae explosions (which heat up the gas, leading to high $\Ratio$).
These evolutionary phases, and hence the lags in time, would result in the spatial offsets
of IR and FUV and keep $\Ratio$ enhanced toward the downstream sides of the spiral arms \citep[see ][]{Egusa:2004yi, Egusa:2009bv, Louie:2013lr}.
The observed $I_{\rm FUV}$ varies by a factor of $\sim 30$ (Figure \ref{fig:ancillary}c),
and, presumably, the CRs follow a similar trend spatially and in flux.
With numerical simulations,
\citet{Penaloza:2017aa, Penaloza:2018aa}, and their series of papers,
showed that $\Ratio$ is elevated in a stronger ISRF and/or at higher CR densities.

The analysis here is based on low-resolution ALMA single-dish data,
but is an important first confirmation of the variations of $\Ratio$.
Obviously, the analysis of one galaxy is limited and a larger sample is called for.
ALMA, with its main interferometer, should be able to resolve $\Ratio$ variations on GMC scales
\citep[][ showed this possibility]{Hirota:2018aa}.
In this paper, we did not consider the possibility that the \textit{majority} of CO emission arises from
the optically-thin CO gas outside GMCs, which could also explain high $\Ratio$.
Such a possibility can be tested when the GMCs are resolved.

 \acknowledgments

This paper makes use of the following ALMA data:
ADS/JAO.ALMA\#2013.1.01161.S, 
ADS/JAO.ALMA\#2015.1.00121.S, 
ADS/JAO.ALMA\#2016.1.00386.S, and 
ADS/JAO.ALMA\#2017.1.00079.S.
ALMA is a partnership of ESO (representing its member states), NSF (USA) and NINS (Japan), together with NRC (Canada), MOST and ASIAA (Taiwan), and KASI (Republic of Korea), in cooperation with the Republic of Chile. The Joint ALMA Observatory is operated by ESO, AUI/NRAO and NAOJ.
The National Radio Astronomy Observatory is a facility of the National Science Foundation operated under cooperative agreement by Associated Universities, Inc..
JK acknowledges support from NSF through grant AST-1812847.
The work of LCH was supported by the National Science Foundation of China (11721303, 11991052) and the National Key R\&D Program of China (2016YFA0400702)
We also thank the anonymous referee.


\facility {ALMA, IRSA, Spitzer, Herschel}
\software{CASA 5.4.1-31 \citep{McMullin:2007aa}}





\end{document}